\documentclass{JHEP3}
\usepackage{amsmath}\allowdisplaybreaks[1]
\usepackage{graphicx}

\newcommand{\Pslash}{\rlap{$\hspace{.34ex}/$}P}
\newcommand{\kslash}{\rlap{$\hspace{.06ex}/$}k}
\newcommand{\Dslash}{\rlap{$\hspace{.38ex}/$}\Delta}
\newcommand{\Nslash}{\rlap{$\hspace{.34ex}/$}N}

\title{Generalized parton correlation functions for a spin-0 hadron}
\author{Stephan Mei{\ss}ner and Klaus Goeke\\
 Institut f{\"u}r Theoretische Physik II, Ruhr-Universit{\"a}t Bochum,\\
 44780 Bochum, Germany\\
 E-mail: \email{stephan.meissner@tp2.rub.de}, \email{klaus.goeke@tp2.rub.de}}
\author{Andreas Metz\\
 Department of Physics, Temple University,\\
 Philadelphia, PA 19122-6082, U.S.A.\\
 E-mail: \email{metza@temple.edu}}
\author{Marc Schlegel\\
 Theory Center, Jefferson Lab, 12000 Jefferson Avenue,\\
 Newport News, VA 23606, U.S.A.\\
 E-mail: \email{schlegel@jlab.org}}
\abstract{The fully unintegrated, off-diagonal quark-quark correlator for a 
spin-0 hadron is parameterized in terms of so-called generalized parton 
correlation functions.
Such objects are of relevance for the phenomenology of certain hard exclusive 
reactions.
In particular, they can be considered as {\it mother distributions} of 
generalized parton distributions on the one hand and transverse momentum 
dependent parton distributions on the other.
Therefore, our study provides new, model-independent insights into the 
recently proposed nontrivial relations between generalized and 
transverse momentum dependent parton distributions.
As a by-product we obtain the first complete classification of generalized 
parton distributions beyond leading twist.}
\preprint{}
\begin{document}

%
%
%
\section{Introduction}
Exploring the partonic substructure of hadrons by means of hard scattering
processes has a long history.
Perhaps the most important reaction in this context is inclusive deep inelastic 
lepton-nucleon scattering providing information on the (ordinary) quark and 
gluon parton distributions (PDFs) of the nucleon.
In the meantime an enormous amount of knowledge about the PDFs has been collected,
especially in the unpolarized case.

Nevertheless, the information contained in PDFs is limited to one dimension in the 
sense that PDFs merely tell us how the parton momenta parallel to the hadron 
momentum are distributed. 
A more complete picture of the parton structure of hadrons is encoded in two other 
types of distributions which are currently subject to intense studies. 
The generalized parton distributions (GPDs)~\cite{Mueller:1998fv} constitute one 
such type.
They can be measured in hard exclusive reactions such as deep virtual Compton 
scattering or hard exclusive meson production (for reviews see, 
e.g., refs.~\cite{Goeke:2001tz,Diehl:2003ny,Belitsky:2005qn,Boffi:2007yc}). 
In particular, when being transformed to the impact parameter space, GPDs contain 
information about the spatial distribution of partons in a plane perpendicular to 
the hadron momentum~\cite{Burkardt:2000za,Ralston:2001xs,Diehl:2002he,Burkardt:2002hr}.
The second kind of generalized functions, the transverse momentum dependent  
parton distributions (TMDs), not only depend on the longitudinal but also on the 
transverse motion of partons inside a hadron.
The TMDs enter the QCD-description of hard semi-inclusive reactions like 
semi-inclusive deep inelastic scattering (SIDIS) or the Drell-Yan (DY) process 
(see, e.g., refs.~\cite{Mulders:1995dh,Bacchetta:2006tn} and the review
articles~\cite{Barone:2001sp,D'Alesio:2007jt}). 

The GPDs and TMDs, {\it a priori}, are considered as independent functions.
However, recently nontrivial relations between these two classes of objects 
have been suggested in the literature \cite{Burkardt:2002ks,Burkardt:2003uw,
Burkardt:2003je,Diehl:2005jf,Burkardt:2005hp,Lu:2006kt,Meissner:2007rx}. 
Of particular interest are the relations between GPDs and the so-called (na\"ive) 
time-reversal odd (T-odd) TMDs like the Sivers 
function~\cite{Sivers:1989cc,Sivers:1990fh} and the Boer-Mulders 
function~\cite{Boer:1997nt}, because they provide an intuitive connection 
between transverse single spin asymmetries observed in different hard 
semi-inclusive processes on the one hand and the distortion of parton distributions 
on the other.
Although many nontrivial relations between GPDs and TMDs were established in 
simple spectator models (see~\cite{Meissner:2007rx} for an overview), no 
model-independent relations have been obtained so far.

The purpose of the present paper now is to investigate the structure of the 
generalized (off-diagonal) quark-quark correlator for a hadron, where for 
simplicity the analysis is restricted to a spin-0 hadron.
This correlator is parameterized in terms of objects which we call generalized 
parton correlation functions (GPCFs).
They depend on the full 4-momentum of the quark and, in addition, contain 
information on the momentum transfer to the hadron.
Both the GPDs as well as the TMDs appear as two different limiting cases of the 
GPCFs.
In other words, this means that the GPCFs can be considered as 
{\it mother distributions} of GPDs and 
TMDs~\cite{Ji:2003ak,Belitsky:2003nz,Belitsky:2005qn}.
The GPCFs also have a direct connection to the so-called Wigner distributions ---
the quantum mechanical analogues of classical phase space distributions --- 
of the hadron-parton system~\cite{Ji:2003ak,Belitsky:2003nz,Belitsky:2005qn}. 

Our motivation for carrying out this study is essentially fourfold:
first, the GPCFs are the most general two-parton correlation functions of 
hadrons.
As such they contain a maximum amount of information about the partonic 
structure of hadrons.
Despite this fact no classification of these objects has been provided in the 
literature so far.
Second, we want to find out which of the GPDs and which of the TMDs arise from 
the same {\it mother distributions}.
Third, we hope, in particular, to get new, model-independent insights into 
the above mentioned nontrivial relations between GPDs and TMDs.
(For a first account on this topic see the conference contribution in 
ref.~\cite{Meissner:2007ez}.)
Fourth, in the case of hard exclusive reactions it is known that, for the kinematics
of current experiments at COMPASS, HERMES, and Jefferson Lab, effects going beyond 
the collinear parton approximation can be quite important 
(see, e.g., refs.~\cite{Vanderhaeghen:1999xj,Diehl:2007hd,Goloskokov:2007nt}), 
though in principle they often can be considered as subleading twist.
(For a related discussion in connection with more inclusive processes 
see~\cite{Collins:2007ph} and references therein.)

The plan of the manuscript is as follows.
In the next section the parameterization of the generalized quark-quark correlator 
in terms of GPCFs is derived.
The results obtained there form the basis for the rest of the paper.
In section \ref{sec3} we consider the so-called generalized transverse momentum dependent
parton distributions (GTMDs), which arise when integrating the GPCFs upon one 
light-cone component of the quark momentum.
The GTMDs are the objects that can directly enter the description of hard exclusive
reactions.
It is worth mentioning that GTMDs for gluons have already been exploited previously 
in order to compute diffractive vector meson~\cite{Martin:1999wb} and 
Higgs production~\cite{Khoze:2000cy}.
In such processes GTMDs appear even at leading order of a twist expansion. 
The TMD-limit and the GPD-limit for the GTMDs are investigated in section \ref{sec4}.
This allows us to obtain the first complete counting of GPDs beyond leading twist.
In particular, we also study which GPDs on the one hand and TMDs on the other have 
the same {\it mother distributions}.
On the basis of our results we are able to investigate the model-independent status 
of possible nontrivial relations between GPDs and TMDs.
Section \ref{sec5} contains the conclusions.
The exact relations between the GPCFs and the GTMDs defined in the manuscript are 
given in appendix~\ref{gtmd_gpcf}, while in appendix~\ref{gtmd_model} our model-independent study is supplemented by the 
calculation of the GTMDs in a simple model for a spin-0 hadron.
%
%
%
\section{Generalized parton correlation functions}
In this section we derive the structure of the generalized, fully-unintegrated
quark-quark correlator for a spin-0 hadron which is defined as
\begin{equation} \label{e:corr_gpcf}
W_{ij}(P,k,\Delta,N;\eta)  =
 \int \frac{d^4 z}{(2 \pi)^4} \, e^{i k \cdot z} \,
 \langle p^{\prime} \, | \, \bar{\psi}_j(-\tfrac{1}{2}z) \,
 {\cal W}(-\tfrac{1}{2}z,\tfrac{1}{2}z\,|\,n) \,
 \psi_i(\tfrac{1}{2}z) \, | \, p \rangle\, .
\end{equation}
The correlator $W$ depends on the average momentum $P = (p+p^\prime)/2$ of the 
initial and final hadron, the momentum transfer $\Delta = p^{\prime} - p$ to 
the hadron, and the average quark momentum $k$.
(For the kinematics we also refer to figure~\ref{f:kinematics}.)
The Wilson line ${\cal W}$ ensures the color gauge invariance of the 
correlator in eq.~(\ref{e:corr_gpcf}) and is running along the path
\begin{equation} \label{e:path}
-\tfrac{1}{2}z \;\to\; -\tfrac{1}{2}z + \infty \cdot n 
\;\to\; \tfrac{1}{2}z + \infty \cdot n \;\to\; \tfrac{1}{2}z \,, 
\end{equation}
with all four points connected by straight lines.
It is now important to realize that the integration contour of the Wilson line 
not only depends on the coordinates of the initial and final points but also 
on the light-cone direction which is opposite to the direction of 
$P$~\cite{Goeke:2003az}.
This induces a dependence on a light-cone vector $n$. 
In fact, instead of using $n$ a rescaled vector $\lambda n$ with some positive 
parameter $\lambda$ could be taken in order to specify the Wilson line.
Therefore, the correlator actually only depends on the vector
\begin{equation} \label{e:direction}
N = \frac{M^2 \, n}{P\cdot n} \,,
\end{equation}
which is invariant under the mentioned rescaling.
For convenience in~(\ref{e:direction}) the hadron mass $M$ is used such that $N$ 
has the same mass dimension as an ordinary 4-momentum.
The parameter $\eta$ in~(\ref{e:corr_gpcf}) is defined through the zeroth component 
of $n$ according to
\begin{equation}
\eta = \textrm{sign}(n_0) \,,
\end{equation} 
which means that we simultaneously treat future-pointing $(\eta = +1)$ and 
past-pointing $(\eta = - 1)$ Wilson lines.
Keeping this dependence is particularly convenient once we make the projection 
of the correlator in~(\ref{e:corr_gpcf}) onto the correlator defining TMDs.
\FIGURE[t]{\includegraphics{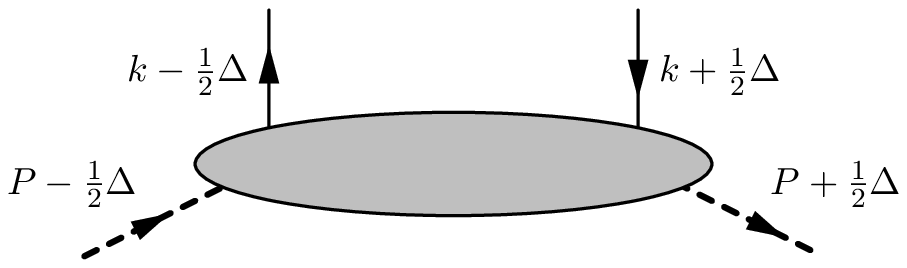}
\caption{Kinematics for GPCFs.}
\label{f:kinematics}}

In order to obtain the parameterization of the correlator in~(\ref{e:corr_gpcf}) 
in terms of GPCFs we make use of the following constraints due to parity, 
hermiticity, and time reversal,
\begin{eqnarray}
W(P,k,\Delta,N;\eta) & = &
\gamma_0 \, W(\bar{P},\bar{k},\bar{\Delta},\bar{N};\eta) \, \gamma_0 \,,
\label{e:w_parity} \\
W^{\dagger}(P,k,\Delta,N;\eta) & = &
\gamma_0 \, W(P,k,-\Delta,N;\eta) \, \gamma_0 \,,
\label{e:w_hermiticity} \\
W^{\ast}(P,k,\Delta,N;\eta) & = & 
\big(-i\gamma_5 C \big) \, W(\bar{P},\bar{k},\bar{\Delta},\bar{N};-\eta) \, 
\big(-i\gamma_5 C \big) \,, 
\label{e:w_timereversal}
\end{eqnarray}
where $\bar{P}^\mu = P_{\mu} = (P^0,-\vec{P})$ etc., while $C$ is the charge 
conjugation matrix. 
It turns out that the general structure of the correlator $W$ can already be obtained 
on the basis of the parity constraint in~(\ref{e:w_parity}).
One ends up with the following 16 linearly independent matrix structures multiplied by 
scalar functions,
\begin{eqnarray} \label{e:w_res}
W(P,k,\Delta,N;\eta)& =& 
  M A_1
+ \Pslash A_2
+ \kslash A_3
+ \Dslash A_4
+ \frac{[\Pslash,\kslash]}{2M} \, A_5
+ \frac{[\Pslash,\Dslash]}{2M} \, A_6
+ \frac{[\kslash,\Dslash]}{2M} \, A_7 
\nonumber\\
& & 
+ \frac{i\varepsilon^{\mu\nu\rho\sigma} 
        \gamma_\mu \gamma_5 P_\nu k_\rho \Delta_\sigma}{M^2} \, A_8
+ \Nslash B_1    
+ \frac{[\Pslash,\Nslash]}{2M} \, B_2
+ \frac{[\kslash,\Nslash]}{2M} \, B_3
\nonumber\\
& &
+ \frac{[\Dslash,\Nslash]}{2M} \, B_4
+ \frac{i\varepsilon^{\mu\nu\rho\sigma} 
        \gamma_\mu \gamma_5 P_\nu k_\rho N_\sigma}{M^2} \, B_5
+ \frac{i\varepsilon^{\mu\nu\rho\sigma} 
        \gamma_\mu \gamma_5 P_\nu \Delta_\rho N_\sigma}{M^2} \, B_6
\nonumber\\
& &
+ \frac{i\varepsilon^{\mu\nu\rho\sigma} 
        \gamma_\mu \gamma_5 k_\nu \Delta_\rho N_\sigma}{M^2} \, B_7
+ \frac{i\varepsilon^{\mu\nu\rho\sigma} 
        \gamma_5 P_\mu k_\nu \Delta_\rho N_\sigma}{M^3} \, B_8 \,.
\end{eqnarray}
Our treatment leading to~(\ref{e:w_res}) is very similar to what has already been 
done for the quark-quark correlator in the case $\Delta = 0$ for a 
spin-0 hadron~\cite{Goeke:2003az,Bacchetta:2004zf} and a 
spin-1/2 hadron~\cite{Goeke:2005hb}.
The functions $A_i$ and $B_i$ are independent and represent the GPCFs.
The various factors of $M$ are introduced in order to assign the same mass dimension 
to all GPCFs.
We emphasize that one has to use the identity 
\begin{equation} 
g^{\alpha \beta} \varepsilon^{\mu \nu \rho \sigma} = 
  g^{\mu \beta} \varepsilon^{\alpha \nu \rho \sigma} 
+ g^{\nu \beta} \varepsilon^{\mu \alpha \rho \sigma} 
+ g^{\rho \beta} \varepsilon^{\mu \nu \alpha \sigma} 
+ g^{\sigma \beta} \varepsilon^{\mu \nu \rho \alpha} \,,
\end{equation}
in order to eliminate redundant terms.
While hermiticity and time-reversal do not affect the general structure of the
correlator $W$, they impose constraints on the GPCFs.
Applying the hermiticity constraint~(\ref{e:w_hermiticity}) to the decomposition 
in~(\ref{e:w_res}) one finds
\begin{equation} \label{e:gpcf_hermiticity}
 X^*(P,k,\Delta,N;\eta) = \pm X(P,k,-\Delta,N;\eta) \,,
\end{equation}
where the plus sign holds for 
$X = A_1$, $A_2$, $A_3$, $A_6$, $A_7$, $A_8$, $B_1$, $B_4$, $B_6$, $B_7$, 
and the minus sign for 
$X = A_4$, $A_5$, $B_2$, $B_3$, $B_5$, $B_8$.
The time-reversal constraint~(\ref{e:w_timereversal}) provides  
\begin{equation} \label{e:gpcf_timereversal}
 X^*(P,k,\Delta,N;\eta) = X(P,k,\Delta,N;-\eta)
\end{equation}
for all $X = A_i,$ $B_i$, relating GPCFs defined with future-pointing Wilson 
lines to those defined with past-pointing lines.
From these considerations it follows that GPCFs, unlike GPDs or TMDs, in general 
are complex-valued functions.
Keeping now in mind that $\eta \in \left\{-1\, ,1\right\}$ and using
eq.~(\ref{e:gpcf_timereversal}) one finds immediately that only the 
imaginary part of the GPCFs depends on $\eta$.
This allows one to write
\begin{equation} \label{e:gpcf_decomp}
 X(P,k,\Delta,N;\eta) = X^{e}(P,k,\Delta,N) + i \, X^{o}(P,k,\Delta,N;\eta) \,,  
\end{equation}
with
\begin{equation} \label{e:gpcf_sign}
X^{o}(P,k,\Delta,N;\eta) = - X^{o}(P,k,\Delta,N;-\eta) \,,   
\end{equation}
where we call $X^{e}$ the T-even and $X^{o}$ the T-odd part of the generic 
GPCF $X$.
The sign reversal of $X^{o}$ in eq.~(\ref{e:gpcf_sign}) when going from 
future-pointing to past-pointing Wilson lines corresponds to the sign reversal 
discussed in ref.~\cite{Collins:2002kn} for T-odd TMDs.

Now we would like to give a first account on the relation between GPCFs on the 
one hand and GPDs as well as TMDs on the other.
To this end we consider the quark-quark correlator $F$ defining GPDs for a spin-0
target, which can be obtained from the correlator $W$ in eq.~(\ref{e:corr_gpcf}) 
by means of the projection
\begin{align} \label{e:corr_gpd}
& F_{ij}(P,x,\Delta,N) =  
 \int dk^- \, d^2\vec{k}_T \, W_{ij}(P,k,\Delta,N;\eta) 
\nonumber \\
& \qquad = \int \frac{dz^-}{2 \pi} \, e^{i k \cdot z} \,
\langle p^{\prime} \, | \, \bar{\psi}_j(-\tfrac{1}{2}z) \,
{\cal W}(-\tfrac{1}{2}z,\tfrac{1}{2}z\,|\,n) \,
\psi_i(\tfrac{1}{2}z) \, | \, p \rangle\, \Big|_{z^+ = \vec{z}_T = 0} \,.
\end{align}
In this formula use is made of light-cone components that are specified 
according to $a^{\pm}=(a^0\pm a^3)/\sqrt{2}$ and $\vec{a}_T = (a^1,a^2)$ 
for a generic 4-vector $a = (a^0,a^1,a^2,a^3)$, where, in particular,
we choose $k^+ = x P^+$.
Note that after integrating upon $k^-$ and $\vec{k}_T$ the dependence on
the parameter $\eta$ drops out. 
As is well-known, in this case we are dealing with a light-cone correlator and 
the two quark fields are just connected by a straight line.
This means that the choice of the contour in~(\ref{e:path}) leads, after projection,
to the appropriate Wilson line for the GPD-correlator.

The correlator $\Phi$ defining TMDs can be extracted from $W$ by putting 
$\Delta = 0$ and integrating out one light-cone component of the quark momentum 
(which we choose to be $k^-$),
\begin{align} \label{e:corr_tmd}
& \Phi_{ij}(P,x,\vec{k}_T,N;\eta) = 
 \int dk^- \, W_{ij}(P,k,0,N;\eta) 
\nonumber \\
& \qquad = \int \frac{dz^- \, d^2 \vec{z}_T}{(2\pi)^3} \, e^{i k \cdot z} \,
 \langle P \, | \, \bar{\psi}_j(-\tfrac{1}{2}z) \,
 {\cal W}(-\tfrac{1}{2}z,\tfrac{1}{2}z\,|\,n) \,
 \psi_i(\tfrac{1}{2}z) \, | \, P \rangle\, \Big|_{z^+ = 0} \,.
\end{align}
Note that for $\Delta = 0$ one has $p = p^{\prime} = P$.
We point out that the path specified in~(\ref{e:path}) also leads to a proper 
Wilson line after taking the 
TMD-limit~\cite{Collins:1981uw,Collins:1999dz,Collins:2000gd,Collins:2002kn,Ji:2002aa,Belitsky:2002sm,Collins:2004nx,Cherednikov:2007tw,Cherednikov:2008ua}.
Since $\Phi$ in eq.~(\ref{e:corr_tmd}) is not a light-cone correlator the dependence 
on the parameter $\eta$ remains.
The case $\eta = +1$ is appropriate for defining TMDs in processes with final state
interactions of the struck quark like SIDIS, while $\eta = -1$ can be used for TMDs 
in DY~\cite{Collins:2002kn}.
It has been emphasized in refs.~\cite{Collins:1981uw,Collins:2003fm,Hautmann:2007uw} that, 
in general, light-like Wilson lines as used in the unintegrated correlators 
in (\ref{e:corr_gpcf}) and (\ref{e:corr_tmd}) lead to divergences.
Such divergences can be avoided, however, by adopting a near light-cone direction.
For the purpose of the present work it is sufficient to note that our general 
reasoning remains valid once a near light-cone direction is used instead of $n$.

It is evident that not only the correlators $F$ and $\Phi$ appear as projections
of the most general two-parton correlator $W$ as outlined above, but also the GPDs 
and the TMDs are projections of certain GPCFs. 
Therefore, GPCFs can be considered as {\it mother distributions}, which actually contain
the maximum amount of information on the two-parton structure of 
hadrons~\cite{Ji:2003ak,Belitsky:2003nz,Belitsky:2005qn}. 
Despite this fact a classification of the GPCFs as given in (\ref{e:w_res}) has 
never been worked out.

%
%
%
\section{Generalized transverse momentum dependent parton distributions}\label{sec3}
The projections in (\ref{e:corr_gpd}) and (\ref{e:corr_tmd}) contain the 
integration upon the minus-component of the quark momentum.
Therefore, it is useful to consider in more detail the correlator
\begin{align} \label{e:corr_gtmd}
& W_{ij}(P,x,\vec{k}_T,\Delta,N;\eta) = 
 \int dk^- \, W_{ij}(P,k,\Delta,N;\eta) 
\nonumber \\
& \qquad = \int \frac{dz^- \, d^2 \vec{z}_T}{(2\pi)^3} \, e^{i k \cdot z} \,
 \langle p^{\prime} \, | \, \bar{\psi}_j(-\tfrac{1}{2}z) \,
 {\cal W}(-\tfrac{1}{2}z,\tfrac{1}{2}z\,|\,n) \,
 \psi_i(\tfrac{1}{2}z) \, | \, p \rangle\, \Big|_{z^+ = 0} \,.
\end{align}
Below the parameterization of this object is given in terms of what we call
generalized transverse momentum dependent parton distributions (GTMDs). 
Of course this result can now be obtained in a straightforward manner on the  
basis of the decomposition in eq.~(\ref{e:w_res}).
From the discussion we have provided so far it is obvious that also the GTMDs, 
like the GPCFs, can be considered as {\it mother distributions} of GPDs and TMDs.
It is the correlator in~(\ref{e:corr_gtmd}) which for instance can enter 
the description of hard exclusive meson production~\cite{Goloskokov:2007nt},
while the corresponding correlator for gluons appears when considering 
diffractive processes in lepton-hadron as well as hadron-hadron 
collisions~\cite{Martin:1999wb,Khoze:2000cy}. 
The question whether or not it appears with a Wilson line as defined 
in~(\ref{e:path}) to our knowledge has never been addressed in the literature 
and requires further investigation that goes beyond the scope of the 
present work.

The matrix of the generalized $k_T$-dependent correlator in (\ref{e:corr_gtmd}) 
is fully specified by means of all possible independent Dirac traces which we 
denote by
\begin{align} \label{e:w_trace}
& W^{[\Gamma]}(P,x,\vec{k}_T,\Delta,N;\eta) = \tfrac{1}{2} \, \text{Tr} 
    \big[ W(P,x,\vec{k}_T,\Delta,N;\eta) \, \Gamma \big]
\nonumber \\
& \qquad = \int \frac{dz^- \, d^2 \vec{z}_T}{2 (2\pi)^3} \, e^{i k \cdot z} \,
 \langle p^{\prime} \, | \, \bar{\psi}(-\tfrac{1}{2}z) \,
 \Gamma \, {\cal W}(-\tfrac{1}{2}z,\tfrac{1}{2}z\,|\,n) \,
 \psi(\tfrac{1}{2}z) \, | \, p \rangle\, \Big|_{z^+ = 0} \,.
\end{align}
In particular, in order to obtain a twist-classification for the GTMDs it is 
convenient to make use of the traces in eq.~(\ref{e:w_trace}).
We choose an infinite momentum frame such that $P$ has a large plus-momentum
and no transverse momentum.
The plus-component of $\Delta$ is expressed through the commonly used 
variable $\xi$.
To be now precise the 4-momenta in (\ref{e:w_res}) are specified according to
\begin{eqnarray}
P & = & \bigg[\, P^+ \, , \, \frac{4M^2 + \vec{\Delta}_T^2}{8(1-\xi^2)P^+} \, , \, 
              \vec{0}_T \, \bigg] \,, \\
k & = & \bigg[\, x P^+ \, ,\, k^- \, , \, \vec{k}_T \, \bigg] \,, \\
\Delta & = & \bigg[\, -2 \xi P^+ \, , \, 
                   \frac{4\xi M^2 + \xi\vec{\Delta}_T^2}{4(1-\xi^2)P^+} \, , \, 
                   \vec{\Delta}_T \, \bigg] \,, \\
 n & = & \bigg[\, 0\, , \, \pm 1 \, , \, \vec{0}_T \, \bigg] \,.
\label{e:lcvec}
\end{eqnarray} 
The vector $n$ in eq.~(\ref{e:lcvec}) is of course not the most general light-cone 
vector.
In particular, it has no transverse component and points opposite to the direction 
of $P$ as already mentioned earlier.
However, if one wants to arrive at an appropriate definition of TMDs for SIDIS and 
DY, there is no freedom left for this vector because it is fixed by the external 
momenta of the processes.

Now we have all the ingredients necessary to write down the final result for 
the traces of the generalized $k_T$-dependent correlator~(\ref{e:corr_gtmd}) in 
terms of GTMDs.
We start with the twist-2 case for which one gets
\begin{eqnarray}
W^{[\gamma^+]} & = & F_1 \,, \vphantom{\frac{1}{1}} 
\label{e:gtmd_1} \\
W^{[\gamma^+ \gamma_5]} & = & 
 \frac{i\varepsilon_T^{ij} k_T^i \Delta_T^j}{M^2} \, \tilde{G}_1 \,, 
\\
W^{[i\sigma^{j+}\gamma_5]} & = & 
 \frac{i\varepsilon_T^{ij} k_T^i}{M} \, H^k_1 
+\frac{i\varepsilon_T^{ij} \Delta_T^i}{M} \, H^\Delta_1 \,. 
\label{e:gtmd_3}
\end{eqnarray}
Here the definitions $\varepsilon^{0123} = 1$ 
and $\varepsilon_T^{ij} = \varepsilon^{-+ij}$ as well as the standard 
notation $\sigma^{\mu\nu} = i [\gamma^{\mu},\gamma^{\nu}] / 2$ are used.
The four complex-valued twist-2 GTMDs $F_1,\, \tilde{G}_1,\, H^k_1,\, H^\Delta_1$
are given by $k^{-}$-integrals of certain linear combinations of the GPCFs 
in~(\ref{e:w_res}), where the explicit relations are listed in appendix~\ref{gtmd_gpcf} 
for all twists.
To shorten the notation the arguments on both sides of the 
eqs.~(\ref{e:gtmd_1})--(\ref{e:gtmd_3}) are omitted.
All GTMDs depend on the set of variables
$(x,\xi,\vec{k}_T^2,\vec{k}_T \cdot \vec{\Delta}_T,\vec{\Delta}_T^2;\eta)$.

In the twist-3 case, characterized through a suppression by one power in $P^{+}$, 
we find
\begin{eqnarray}
W^{[1]} & = & \frac{M}{P^+}
 \bigg[ E_2 \bigg] \,,
\\
W^{[\gamma_5]} & = & \frac{M}{P^+}
 \bigg[ \frac{i\varepsilon_T^{ij} k_T^i \Delta_T^j}{M^2} \, \tilde{E}_2 \bigg] \,,
\\
W^{[\gamma^j]} & = & \frac{M}{P^+}
 \bigg[ \frac{k_T^j}{M} \, F^k_2 + \frac{\Delta_T^j}{M} \, F^\Delta_2 \bigg] \,,
\\
W^{[\gamma^j\gamma_5]} & = & \frac{M}{P^+}
 \bigg[ \frac{i\varepsilon_T^{ij} k_T^i}{M} \, G^k_2
       +\frac{i\varepsilon_T^{ij} \Delta_T^i}{M} \, G^\Delta_2 \bigg] \,,
\\
W^{[i\sigma^{ij}\gamma_5]} & = & \frac{M}{P^+}
 \bigg[ i\varepsilon_T^{ij} H_2 \bigg] \,,
\\
W^{[i\sigma^{+-}\gamma_5]} & = & \frac{M}{P^+}
 \bigg[ \frac{i\varepsilon_T^{ij} k_T^i \Delta_T^j}{M^2} \, \tilde{H}_2 \bigg] \,.
\end{eqnarray}
The twist-4 result, which is basically a copy of the twist-2 case, reads
\begin{eqnarray}
W^{[\gamma^-]} & = & \frac{M^2}{(P^+)^2}
 \bigg[ F_3 \bigg] \,,
\\
W^{[\gamma^- \gamma_5]} & = & \frac{M^2}{(P^+)^2}
 \bigg[ \frac{i\varepsilon_T^{ij} k_T^i \Delta_T^j}{M^2} \, \tilde{G}_3 \bigg] \,,
\\
W^{[i\sigma^{j-}\gamma_5]} & = & \frac{M^2}{(P^+)^2}
 \bigg[ \frac{i\varepsilon_T^{ij} k_T^i}{M} \, H^k_3
       +\frac{i\varepsilon_T^{ij} \Delta_T^i}{M} \, H^\Delta_3 \bigg] \,.
\label{e:gtmd_12}
\end{eqnarray}
The twist-4 case is of course at most of academic interest but is included for 
completeness. 

Like in the case of the GPCFs we also consider the implications of hermiticity and
time-reversal on the GTMDs.
Hermiticity leads to
\begin{equation} \label{e:gtmd_hermiticity}
 X^*(x,\xi,\vec{k}_T^2,\vec{k}_T \cdot \vec{\Delta}_T,\vec{\Delta}_T^2;\eta)
 =\pm X(x,-\xi,\vec{k}_T^2,-\vec{k}_T \cdot \vec{\Delta}_T,\vec{\Delta}_T^2;\eta) \,,
\end{equation}
with a plus sign for $X = E_2$, $F_1$, $F_2^k$, $F_3$, $\tilde{G}_1$, 
$G_2^{\Delta}$, $\tilde{G}_3$, $H_1^{\Delta}$, $\tilde{H}_2$, $H_3^{\Delta}$,
and the minus sign for $X = \tilde{E}_2$, $F_2^{\Delta}$, $G_2^k$, $H_1^k$, 
$H_2$, $H_3^k$.
These results are a direct consequence of~(\ref{e:gpcf_hermiticity}) and the 
relations in eqs.~(\ref{e:gtmd_gpcf_1})--(\ref{e:gtmd_gpcf_16}).
On the basis of~(\ref{e:gpcf_timereversal}) one obtains from time-reversal
\begin{equation}
X^*(x,\xi,\vec{k}_T^2,\vec{k}_T \cdot \vec{\Delta}_T,\vec{\Delta}_T^2;\eta) = 
X(x,\xi,\vec{k}_T^2,\vec{k}_T \cdot \vec{\Delta}_T,\vec{\Delta}_T^2;-\eta) \,. 
\label{linear}
\end{equation}
for all GTMDs $X$.
This means, in particular, that we can carry over eqs.~(\ref{e:gpcf_decomp}) 
and (\ref{e:gpcf_sign}) to the GTMD case and write
\begin{equation}
 X(x,\xi,\vec{k}_T^2,\vec{k}_T \cdot \vec{\Delta}_T,\vec{\Delta}_T^2;\eta)
 = X^{e}(x,\xi,\vec{k}_T^2,\vec{k}_T \cdot \vec{\Delta}_T,\vec{\Delta}_T^2) 
  +  i \, X^{o}(x,\xi,\vec{k}_T^2,\vec{k}_T \cdot \vec{\Delta}_T,\vec{\Delta}_T^2;\eta) \,, 
\end{equation}
with the real valued functions $X^{e}$ and $X^{o}$ respectively representing the 
real and imaginary part of the GTMD $X$.
Only the T-odd part $X^{o}$ depends on the sign of $\eta$ according to 
\begin{equation}
X^{o}(x,\xi,\vec{k}_T^2,\vec{k}_T \cdot \vec{\Delta}_T,\vec{\Delta}_T^2;\eta) = - 
X^{o}(x,\xi,\vec{k}_T^2,\vec{k}_T \cdot \vec{\Delta}_T,\vec{\Delta}_T^2;-\eta) \,, 
\end{equation}
i.e., the imaginary parts of GTMDs defined with future-pointing and past-pointing 
Wilson lines have a reversed sign.

In order to give an estimate we have calculated the GTMDs in a simple spectator 
model of the pion.
The results are presented in appendix~\ref{gtmd_model}.
Our treatment is restricted to lowest order in perturbation theory.
To this order all T-odd parts of the GTMDs vanish --- a feature which is also 
well-known from spectator model calculations of T-odd TMDs.
All the results listed in eqs.~(\ref{e:gtmd_model_1})--(\ref{e:gtmd_model_16}) 
are in accordance with the hermiticity constraint~(\ref{e:gtmd_hermiticity}).

%
%
%
\section{Projection of GTMDs onto TMDs and GPDs}\label{sec4}
In this section we consider the generalized $k_T$-dependent correlator in 
eq.~(\ref{e:corr_gtmd}) for the specific TMD-kinematics and the GPD-kinematics.
This procedure provides the relations between the {\it mother distributions} (GTMDs)
on the one hand and the TMDs as well as GPDs on the other.
On the basis of these results one can check whether there exists  
model-independent support for possible nontrivial relations between GPDs 
and TMDs.

\subsection{TMD-limit}
We start with the TMD-limit corresponding to a vanishing momentum transfer 
$\Delta = 0$.
In this limit exactly half of the real-valued distributions vanish because they
are odd as function of $\Delta$ due to the hermiticity 
constraint~(\ref{e:gtmd_hermiticity}):
$E_2^{o}$, $\tilde{E}_2^{e}$, $F_1^{o}$, $F_2^{k,o}$, 
$F_2^{\Delta,e}$, $F_3^{o}$, $\tilde{G}_1^{o}$, $G_2^{k,e}$, 
$G_2^{\Delta,o}$, $\tilde{G}_3^{o}$, $H_1^{k,e}$, $H_1^{\Delta,o}$, 
$H_2^{e}$, $\tilde{H}_2^{o}$, $H_3^{k,e}$, $H_3^{\Delta,o}$.
In addition, the distributions 
$\tilde{E}_2^{o}$, $F_2^{\Delta,o}$, $\tilde{G}_1^{e}$, $G_2^{\Delta,e}$, 
$\tilde{G}_3^{e}$, $H_1^{\Delta,e}$, $\tilde{H}_2^{e}$, $H_3^{\Delta,e}$ 
do not appear in the correlator any more, because they are multiplied by a 
coefficient which is linear in $\Delta$.
Therefore, in the TMD-limit only the following eight (four T-even and four T-odd)
distributions survive:
$E_2^{e}$, $F_1^{e}$, $F_2^{k,e}$, $F_3^{e}$, 
$G_2^{k,o}$, $H_1^{k,o}$, $H_2^{o}$, $H_3^{k,o}$. 
The complete list of TMDs for a spin-1/2 hadron has been given in 
ref.~\cite{Goeke:2005hb}\footnote{Note that the {\it l.h.s.} in eqs.~(16) and~(25) 
of~\cite{Goeke:2005hb} should read $\Phi^{[i\sigma^{i+}\gamma_5]}$ and 
$\Phi^{[i\sigma^{i-}\gamma_5]}$, respectively.} 
(see also the review article~\cite{Bacchetta:2006tn}).
Considering in~\cite{Goeke:2005hb} the limit of a spinless target one ends up 
with eight TMDs which agrees with the number of GTMDs in the limit $\Delta = 0$.
Using for the TMDs the notation of~\cite{Goeke:2005hb} one finds the following 
explicit relations between the TMDs and the GTMDs:
\begin{eqnarray}
 f_1(x,\vec{k}_T^2) & = & 
 F_1^{e}(x,0,\vec{k}_T^2,0,0) \,,
\label{e:tmd_gtmd_1} \\
 h^\bot_1(x,\vec{k}_T^2;\eta) & = &
 H_1^{k,o}(x,0,\vec{k}_T^2,0,0;\eta) \,,
\label{e:tmd_gtmd_2} \\
 e(x,\vec{k}_T^2) & = &
 E_2^{e}(x,0,\vec{k}_T^2,0,0) \,,
\\
 f^\bot(x,\vec{k}_T^2) & = &
 F_2^{k,e}(x,0,\vec{k}_T^2,0,0) \,,
\\
 g^\bot(x,\vec{k}_T^2;\eta) & = &
 G_2^{k,o}(x,0,\vec{k}_T^2,0,0;\eta) \,,
\\
 h(x,\vec{k}_T^2;\eta) & = & 
 H_2^{o}(x,0,\vec{k}_T^2,0,0;\eta) \,,
\\
 f_3(x,\vec{k}_T^2) & = &
 F_3^{e}(x,0,\vec{k}_T^2,0,0) \,,
\\
 h^\bot_3(x,\vec{k}_T^2;\eta) & = &
 H_3^{k,o}(x,0,\vec{k}_T^2,0,0;\eta) \,.
\end{eqnarray}
These results are obtained by means of 
eqs.~(\ref{e:corr_tmd}),~(\ref{e:corr_gtmd}), and~(\ref{e:w_trace}), together with 
the explicit expressions for the traces of the TMD-correlator $\Phi$ in terms of 
TMDs as given in~\cite{Goeke:2005hb}, and the traces of the GTMD-correlator 
in~(\ref{e:gtmd_1})--(\ref{e:gtmd_12}).
The four TMDs $h_1^\bot$, $g^\bot$, $h$, $h_3^\bot$ are T-odd and are related to 
T-odd GTMDs.

\subsection{GPD-limit}

In a second step we concentrate on the GPD-limit which appears when integrating 
upon the transverse parton momentum $\vec{k}_T$. 
As already discussed after eq.~(\ref{e:corr_gpd}) the dependence on $\eta$ drops 
out in this case which implies, in particular, that all effects of T-odd GTMDs 
disappear.

In the literature only the twist-2 and the chiral-even twist-3 GPDs for a spin-0
target have been introduced~\cite{Anikin:2000em,Belitsky:2005qn}.
Therefore, we give here for the first time a complete list of GPDs for all 
twists.
The GPDs parameterize the Dirac traces $F^{[\Gamma]}$ of the GPD-correlator 
in~(\ref{e:corr_gpd}).
One finds two nonvanishing traces for twist-2, four traces for twist-3, and two 
traces for twist-4.
To be explicit the GPDs can be defined according to
\begin{eqnarray}
F^{[\gamma^+]} & = & F_1^{\pi}(x,\xi,t) \,, \vphantom{\frac{1}{1}} 
\label{e:gpd_1} \\
F^{[i\sigma^{j+}\gamma_5]} & = & 
 \frac{i\varepsilon_T^{ij} \Delta_T^i}{M} \, H_1^{\pi}(x,\xi,t) \,, 
\\
F^{[1]} & = & \frac{M}{P^+}
 \bigg[ E_2^{\pi}(x,\xi,t) \bigg] \,,
\\
F^{[\gamma^j]} & = & \frac{M}{P^+}
 \bigg[ \frac{\Delta_T^j}{M} \, F_2^{\pi}(x,\xi,t) \bigg] \,,
\\
F^{[\gamma^j\gamma_5]} & = & \frac{M}{P^+}
 \bigg[ \frac{i\varepsilon_T^{ij} \Delta_T^i}{M} \, G_2^{\pi}(x,\xi,t) \bigg] \,,
\\
F^{[i\sigma^{ij}\gamma_5]} & = & \frac{M}{P^+}
 \bigg[ i\varepsilon_T^{ij} H_2^{\pi}(x,\xi,t) \bigg] \,,
\\
F^{[\gamma^-]} & = & \frac{M^2}{(P^+)^2}
 \bigg[ F_3^{\pi}(x,\xi,t) \bigg] \,,
\\
F^{[i\sigma^{j-}\gamma_5]} & = & \frac{M^2}{(P^+)^2}
 \bigg[ \frac{i\varepsilon_T^{ij} \Delta_T^i}{M} \, H_3^{\pi}(x,\xi,t) \bigg] \,,
\label{e:gpd_8}
\end{eqnarray}
where $t = \Delta^2$.
The structure of the traces in~(\ref{e:gpd_1})--(\ref{e:gpd_8}) follows readily 
from eqs.~(\ref{e:gtmd_1})--(\ref{e:gtmd_12}) if one keeps in mind that after 
integrating upon $\vec{k}_T$ the only transverse vector left is $\vec{\Delta}_T$.
Altogether there exist eight GPDs corresponding to the number of TMDs for a 
spin-0 hadron.
The four GPDs $F_1^{\pi}$, $F_2^{\pi}$, $G_2^{\pi}$, $F_3^{\pi}$ are chiral-even, 
while the remaining ones are chiral-odd.
In the next subsection we consider in more detail the GPD $H_1^{\pi}$ which is
related to the object $\bar{E}_T$ used in~\cite{Burkardt:2007xm} according to
$H_1^{\pi} = - \bar{E}_T$.
The twist-3 GPDs $F_2^{\pi}$ and $G_2^{\pi}$ are related to the functions
$H_3$ and $H_A$ that were introduced in ref.~\cite{Anikin:2000em}.
At this point it is also worthwhile to notice that, in general, measuring 
chiral-odd GPDs would be a demanding task.
For instance in the case of hard exclusive production of a single meson one 
would have to consider subleading twist observables in order to get access to 
these objects~\cite{Diehl:1998pd,Collins:1999un}.
Alternatively one could resort to relatively complicated processes like 
diffractive electroproduction of two mesons~\cite{Ivanov:2002jj}.
On the other hand chiral-odd GPDs can well be investigated using 
lattice QCD~\cite{Brommel:2007xd} or models of the nonperturbative strong 
interaction.

It is now straightforward to write down the following expressions for the GPDs 
in terms of $k_T$-integrals of GTMDs:
\begin{eqnarray} 
F_1^{\pi}(x,\xi,t) & = & \int d^2\vec{k}_T \, 
 F_1^{e} \,,
\label{e:gpd_gtmd_1} \\
H_1^{\pi}(x,\xi,t) & = & \int d^2\vec{k}_T \, 
\bigg[ \frac{\vec{k}_T \cdot \vec{\Delta}_T}{\vec{\Delta}_T^2} \, H_1^{k,e} +
 H_1^{\Delta,e} \bigg] \,,
\label{e:gpd_gtmd_2} \\
E_2^{\pi}(x,\xi,t) & = & \int d^2\vec{k}_T \, 
 E_2^{e} \,,
\\
F_2^{\pi}(x,\xi,t) & = & \int d^2\vec{k}_T \, 
\bigg[ \frac{\vec{k}_T \cdot \vec{\Delta}_T}{\vec{\Delta}_T^2} \, F_2^{k,e} +
 F_2^{\Delta,e} \bigg] \,,
\\
G_2^{\pi}(x,\xi,t) & = & \int d^2\vec{k}_T \, 
\bigg[ \frac{\vec{k}_T \cdot \vec{\Delta}_T}{\vec{\Delta}_T^2} \, G_2^{k,e} +
 G_2^{\Delta,e} \bigg] \,,
\\
H_2^{\pi}(x,\xi,t) & = & \int d^2\vec{k}_T \, 
 H_2^{e} \,,
\\
F_3^{\pi}(x,\xi,t) & = & \int d^2\vec{k}_T \, 
 F_3^{e} \,,
\\
H_3^{\pi}(x,\xi,t) & = & \int d^2\vec{k}_T \, 
\bigg[ \frac{\vec{k}_T \cdot \vec{\Delta}_T}{\vec{\Delta}_T^2} \,  H_3^{k,e} +
 H_3^{\Delta,e} \bigg] \,.
\label{e:gpd_gtmd_8}
\end{eqnarray}
Note that in eqs.~(\ref{e:gpd_gtmd_1})--(\ref{e:gpd_gtmd_8}) the limit 
$\vec{\Delta}_T \to 0$ can be performed without encountering a singularity 
because of
\begin{equation}
\int d^2\vec{k}_T \, k_T^i \, 
X(x,\xi,\vec{k}_T^2,\vec{k}_T \cdot \vec{\Delta}_T,\vec{\Delta}_T^2;\eta) 
\propto \Delta_T^i \,,
\end{equation}
which holds for any GTMD $X$. 
The hermiticity constraint~(\ref{e:gtmd_hermiticity}) for the GTMDs, in 
combination with the relations~(\ref{e:gpd_gtmd_1})--(\ref{e:gpd_gtmd_8}), 
determines the symmetry behavior of the GPDs under the transformation
$\xi \to - \xi$.
One finds that the six GPDs $E_2^{\pi}$, $F_1^{\pi}$, $F_3^{\pi}$, $G_2^{\pi}$, 
$H_1^{\pi}$, $H_3^{\pi}$ are even functions in $\xi$, while
$F_2^{\pi}$ and $H_2^{\pi}$ are odd in $\xi$.
This implies
\begin{equation} \label{e:gpd_symm}
F_2^{\pi}(x,0,t) = 0 \,, \qquad
H_2^{\pi}(x,0,t) = 0 \,. 
\end{equation}
In the following subsection we will make use of~(\ref{e:gpd_symm}).

\subsection{Relations between GPDs and TMDs}

Having established the precise connection of the GPDs and TMDs with their 
respective {\it mother distributions} we are now in a position to search for 
possible model-independent relations between GPDs and TMDs.
From~(\ref{e:gpd_gtmd_1}) and~(\ref{e:tmd_gtmd_1}) it is obvious that the 
GPD $F_1^{\pi}$ and the TMD $f_1$ can be related since both functions are 
projections of the GTMD $F_1^e$.
With an analogous reasoning also one relation between twist-3 and one 
between twist-4 distributions can be obtained leading altogether to  
\begin{eqnarray}
F_1^{\pi}(x,0,0) & = & \int d^2\vec{k}_T \, F_1^e(x,0,\vec{k}^2_T,0,0)
= \int d^2 \vec{k}_T \, f_1(x,\vec{k}^2_T) = f_1(x) \,,
\\
E_2^{\pi}(x,0,0) & = & \int d^2\vec{k}_T \, E_2^e(x,0,\vec{k}^2_T,0,0)
= \int d^2 \vec{k}_T \, e(x,\vec{k}^2_T) = e(x) \,,
\\
F_3^{\pi}(x,0,0) & = & \int d^2\vec{k}_T \, F_3^e(x,0,\vec{k}^2_T,0,0)
= \int d^2 \vec{k}_T \, f_3(x,\vec{k}^2_T) = f_3(x) \,.
\end{eqnarray}
These formulas have the same status as the (trivial) model-independent 
relations between GPDs and TMDs which are known for certain twist-2 quark 
and gluon distributions of the nucleon (called relations of first type 
in ref.~\cite{Meissner:2007rx}).

However, also nontrivial relations between GPDs and TMDs have been suggested 
in the literature~\cite{Burkardt:2002ks,Burkardt:2003uw,Burkardt:2003je,Diehl:2005jf,Burkardt:2005hp,Lu:2006kt,Meissner:2007rx} in the case of a spin-1/2 target.
So far these relations have only been established in low-order calculations
in the framework of simple spectator models.
Our GTMD-analysis can now shed light on the question if model-independent 
nontrivial relations exist.
\FIGURE[t]{\includegraphics{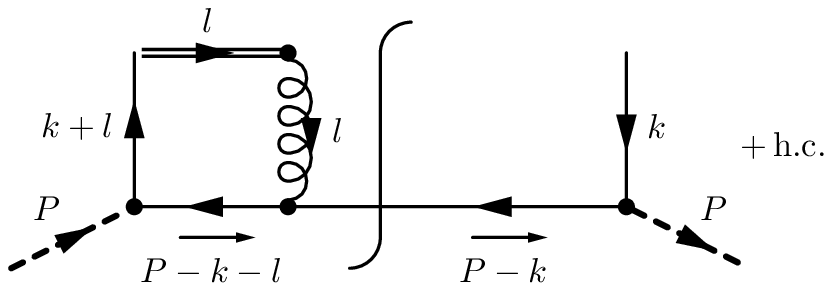}
\caption{Lowest nontrivial order diagram for T-odd TMDs in the spectator model 
for a pion.
The Hermitian conjugate diagram (h.c.) is not shown.
The eikonal propagator arising from the Wilson line in the operator definition
of TMDs is indicated by a double line.}
\label{f:bm}}

In the case of a nucleon target it was shown that for instance the Boer-Mulders
function $h_1^\perp$ has a (model-dependent) nontrivial relation to a certain 
linear combination of chiral-odd 
GPDs~\cite{Diehl:2005jf,Burkardt:2005hp,Meissner:2007rx}.
This result suggests for a spin-$0$ target a relation between $h_1^\perp$ and 
$H_1^{\pi}$.
We can investigate this issue using the simple spectator model of the pion discussed
in appendix~\ref{gtmd_model}.
The GPD $H_1^{\pi}$ is obtained by means of eq.~(\ref{e:gpd_gtmd_2}) and the results 
for the GTMDs in~(\ref{e:gtmd_model_11}) and~(\ref{e:gtmd_model_12}).
One finds
\begin{eqnarray}
H_1^{\pi}(x,0,-\tfrac{\vec{\Delta}_T^2}{(1-x)^2}) & = &
- \frac{g_{\pi}^2 \, (1-x)}{(2\pi)^3} \int d^2\vec{k}_T \, 
\frac{m \, M}{[\vec{k}_T^2 + \tilde{M}^2(x)]
         [(\vec{k}_T + \vec{\Delta}_T)^2 + \tilde{M}^2(x)]} \,,
\nonumber \\
&& \textrm{with} \quad 
\tilde{M}^2(x) = m^2 - x(1-x) M^2 \,.
\vphantom{\frac{1}{1}}
\label{e:gpd_psm}
\end{eqnarray}
Note that, in general, relations between GPDs and TMDs always invoke GPDs that
are evaluated at the specific kinematical point $\xi = 0$~\cite{Meissner:2007rx}.
A nonzero Boer-Mulders function may be generated by including contributions
from the Wilson line in the TMD correlator.
To lowest nontrivial order one has to consider the diagram shown in 
figure~\ref{f:bm}, where a gluon is exchanged between the high-energy (eikonalized)
quark and the spectator parton.
(For related treatments of the Boer-Mulders function of the nucleon we refer
to~\cite{Goldstein:2002vv,Boer:2002ju,Gamberg:2003ey,Bacchetta:2003rz,Gamberg:2007wm}.)
In the case of valence quarks and antiquarks of a neutral pion we 
find\footnote{A calculation of $h_1^\perp$ of the pion using the pseudo-scalar 
model defined in appendix~\ref{gtmd_model} has already been performed in ref.~\cite{Lu:2004hu}, 
but the result quoted in that paper does not fully agree with ours. 
Both results differ, in particular, by a factor $\tfrac{1}{2}(1-x)$.}
\begin{equation}
h_1^\perp(x,\vec{k}_T^2;\eta) = - \eta \, \frac{2 g_{\pi}^2 \, g_s^2}{3 (2\pi)^4} \,
 \frac{m \, M}{\vec{k}_T^2 \, (\vec{k}_T^2 + \tilde{M}^2(x))} \,
\ln \bigg( \frac{\vec{k}_T^2 + \tilde{M}^2(x)}{\tilde{M}^2(x)}\bigg) \,,
\label{e:tmd_psm}
\end{equation}
with $g_s$ denoting the strong coupling.
One has to multiply the expression in~(\ref{e:tmd_psm}) by two in order to get
the valence Boer-Mulders functions of charged pions.
Obviously $h_1^\perp$ is negative which agrees with previous 
expectations~\cite{Burkardt:2007xm}.
Using straightforward manipulations one can now show that the distributions 
in~(\ref{e:gpd_psm}) and~(\ref{e:tmd_psm}) are related according to 
\begin{equation}
\frac{2 g_s^2}{3 (2\pi)^2 \, (1-x)} 
\int d^2\vec{\Delta}_T \, H_1^{\pi}(x,0,-\tfrac{\vec{\Delta}_T^2}{(1-x)^2}) =
\eta \int d^2 \vec{k}_T \, \frac{\vec{k}_T^2}{2M^2} \, 
h_1^\perp(x,\vec{k}_T^2;\eta) \,.
\label{e:rel_psm}
\end{equation}
This equation is of the form of the nontrivial relation of second type
presented in eq.~(106) of ref.~\cite{Meissner:2007rx}. 
We point out that there exist different ways of writing the relation between 
$H_1^{\pi}$ and $h_1$.
For instance one may transform the GPD into impact parameter space which leads 
to an interesting physical interpretation of the 
relation~\cite{Burkardt:2002ks,Burkardt:2003je,Diehl:2005jf,Meissner:2007rx}.
Moreover, relations invoking different $\Delta_T$-moments of GPDs and $k_T$-moments 
of TMDs can also be established~\cite{Lu:2006kt,Meissner:2007rx}.
For the purpose of our discussion here it is sufficient to notice that in the 
framework of a simple spectator model of the pion a nontrivial relation between 
$H_1^\pi$ and $h_1^\perp$ holds.

On the other hand, the GTMD-analysis does not support a model-independent status 
of such a relation because, according to~(\ref{e:gpd_gtmd_2}) 
and~(\ref{e:tmd_gtmd_2}), $H_1^\pi$ and $h_1^\perp$ have different, independent 
{\it mother distributions}.
One may wonder if the specific kinematical point $\xi = 0$ and the use of
moments as taken in~(\ref{e:rel_psm}) spoils the argument.
However, it turns out that the contributions of all the involved GTMDs
survive these operations.
Unless, for some reason, the involved GTMDs are subject to further constraints 
one has to conclude that there cannot exist a model-independent relation between 
$H_1^\pi$ and $h_1^\perp$.
This conclusion is in accordance with the observation made 
in~\cite{Meissner:2007rx} that nontrivial relations of second type are 
likely to even break down in spectator models once higher order contributions 
are taken into account.
Therefore, one has to attribute the relation to the simplicity of the used
model.
Nevertheless, it may well be that numerically the model-dependent nontrivial
relation works reasonably well when comparing to experimental data.
In fact such a case is already known for distributions of the nucleon, 
namely the relation between the Sivers function $f_{1T}^\perp$ and the 
GPD $E$~\cite{Burkardt:2002ks,Burkardt:2003uw,Burkardt:2003je,Meissner:2007rx}.

In order to continue we now note that according to eq.~(\ref{e:gpd_symm}) the 
GPDs $F_2^\pi$ and $H_2^\pi$ vanish for $\xi = 0$.
Therefore, they cannot be related to TMDs.
This means we are just left with the GPDs $G_2^\pi$ and $H_3^\pi$.
Since their potential counterparts $g^\perp$ and $h_3^\perp$ on the TMD side 
are again related to different GTMDs also in that case no model-independent 
relation exists.
Eventually, we mention that for the subleading twist functions we have not 
explored whether any model-dependent nontrivial relation can be obtained.
%
%
%
\section{Conclusions}\label{sec5}
In summary, we have derived the structure of the fully unintegrated, off-diagonal 
quark-quark correlator for a spin-0 hadron. 
This object, which contains the most general information on the two-parton 
structure of a hadron, has been parameterized in terms of so-called generalized
parton correlation functions (GPCFs).
Integrating the GPCFs upon a light-cone component of the quark momentum one ends
up with entities which we called generalized transverse momentum dependent parton 
distributions (GTMDs).
In general, GTMDs can be of direct relevance for the phenomenology of various hard 
(diffractive) processes 
(see, e.g., refs.~\cite{Martin:1999wb,Khoze:2000cy,Goloskokov:2007nt}).
Our analysis shows that both the GPCFs and the GTMDs in general are complex-valued 
functions.
This is different from the (simpler) forward parton distributions, GPDs, and TMDs 
all of which are real.

Suitable projections of GTMDs lead to GPDs on the one hand and TMDs on the other.
Therefore, GTMDs can be considered as {\it mother distributions} of GPDs and 
TMDs~\cite{Ji:2003ak,Belitsky:2003nz,Belitsky:2005qn}. 
To study these two limiting cases of GTMDs was the main motivation of the present 
work.
One outcome was the first complete classification of GPDs beyond leading twist.
Most importantly, we were able to determine which of the GPDs and TMDs have the 
same {\it mother distributions} allowing us to explore whether model-independent 
relations between GPDs and TMDs can be established.
For a spin-0 hadron one ends up with three such model-independent relations.
Actually, these cases can be considered as trivial ones because the respective 
GPDs and TMDs also have a relation to the same forward parton distributions 
(see also ref.~\cite{Meissner:2007rx}).
Our main interest was to investigate nontrivial relations between GPDs and TMDs 
which have been obtained in simple spectator models and extensively discussed 
in the recent 
literature~\cite{Burkardt:2002ks,Burkardt:2003uw,Burkardt:2003je,Diehl:2005jf,Burkardt:2005hp,Lu:2006kt,Meissner:2007rx}.
We have elaborated in some detail on a possible relation between the (twist-2)
Boer-Mulders function $h_1^\perp$ of a pion and a chiral-odd GPD 
(denoted by $H_1^\pi$). 
While a (nontrivial) relation can be found in a low order calculation using a 
simple spectator model of the pion, the GTMD-analysis shows that such a relation
cannot have a model-independent status because $h_1^\perp$ and $H_1^\pi$ are 
related to different, independent {\it mother distributions}.
This finding agrees with ref.~\cite{Meissner:2007rx} where it has been argued that 
nontrivial relations between GPDs and TMDs are likely to break down even in 
spectator models if the parton distributions are evaluated to higher order in 
perturbation theory.
Altogether we found that only the three mentioned trivial relations can be 
model-independent.
We emphasize that this conclusion does not tell anything about the numerical 
violation of possible nontrivial relations between GPDs and TMDs. 

The present work should be extended in various directions.
First of all, it would be worthwhile to repeat our analysis for the more
interesting but at the same time also more complicated case of a nucleon 
target.
Moreover, one should focus on the phenomenology of GPCFs and GTMDs.
For instance, in this context it has to be clarified if the GTMDs in physical 
processes appear with the Wilson line as defined in our work.
To our knowledge previous articles using GTMDs did not address this important 
question.
In addition, if one wants to study diffractive reactions, gluon GTMDs rather 
than quark GTMDs are relevant in a first place.
We hope to return to these topics in future work.
%
%
%
\acknowledgments 
We are grateful to H.~Avakian and M.~Diehl for useful discussions. 
The work has partially been supported by the Verbundforschung ``Hadronen und Kerne''
of the BMBF and by the Deutsche Forschungsgemeinschaft (DFG).
\\[0.3cm]
\noindent
\textbf{Notice:} Authored by Jefferson Science Associates, LLC under U.S. DOE 
Contract No. DE-AC05-06OR23177. 
The U.S. Government retains a non-exclusive, paid-up, irrevocable, world-wide license 
to publish or reproduce this manuscript for U.S. Government purposes. 

%
%
%
\appendix
\section{Relations between GTMDs and GPCFs} 
\label{gtmd_gpcf}
Here the explicit relations between the GTMDs in 
eqs.~(\ref{e:gtmd_1})--(\ref{e:gtmd_12}) and the GPCFs in eq.~(\ref{e:w_res}) 
are listed.
For brevity we leave out the arguments of the functions.
Straightforward calculation leads to the results
\begin{eqnarray}
 E_2
 &=& 2P^+ \, \int dk^- \, \bigg[ A_1 \bigg] \,, \label{e:gtmd_gpcf_1} \\
 \tilde{E}_2
 &=& 2P^+ \, \int dk^- \, \bigg[ B_8 \bigg] \,,\\
 F_1
 &=& 2P^+ \, \int dk^- \, \bigg[ A_2 + x A_3 - 2\xi A_4 \bigg] \,, \\
 F_2^k
 &=& 2P^+ \, \int dk^- \, \bigg[ A_3 \bigg] \,, \\
 F_2^\Delta
 &=& 2P^+ \, \int dk^- \, \bigg[ A_4 \bigg] \,, \\
 F_3
 &=& 2P^+ \, \int dk^- \, \bigg[ \frac{P^2}{2M^2} \, \big( A_2 - x A_3 + 2\xi A_4 \big)
     + \frac{P \cdot k}{M^2} \, A_3 + B_1 \bigg]  \,, \\
 \tilde{G}_1
 &=& 2P^+ \, \int dk^- \, \bigg[ A_8 \bigg] \,, \\
 G_2^k
 &=& 2P^+ \, \int dk^- \, \bigg[ \frac{2\xi P^2}{M^2} \, A_8 + B_5 + 2\xi B_7 \bigg] \,, \\
 G_2^\Delta
 &=& 2P^+ \, \int dk^- \, \bigg[ \frac{x P^2 - P \cdot k}{M^2} \, A_8 + B_6 + x B_7 \bigg] \,, \\
 \tilde{G}_3
 &=& 2P^+ \, \int dk^- \, \bigg[ -\frac{P^2}{2M^2} \, A_8 - B_7 \bigg] \,, \\
 H_1^k
 &=& 2P^+ \, \int dk^- \, \bigg[ - A_5 - 2\xi A_7 \bigg] \,, \\
 H_1^\Delta
 &=& 2P^+ \, \int dk^- \, \bigg[ - A_6 - x A_7 \bigg] \,, \\
 H_2
 &=& 2P^+ \, \int dk^- \, \bigg[ \frac{P^2}{M^2} \, \big( - x A_5 + 2\xi A_6 \big)
     + \frac{P \cdot k}{M^2} \, \big( A_5 + 2\xi A_7 \big) \nonumber\\*
 & & \hspace{14.15ex} + B_2 + x B_3 - 2\xi B_4 \bigg] \,, \\
 \tilde{H}_2
 &=& 2P^+ \, \int dk^- \, \bigg[ - A_7 \bigg] \,, \\
 H_3^k
 &=& 2P^+ \, \int dk^- \, \bigg[ \frac{P^2}{2 M^2} \, \big( A_5 - 2\xi A_7 \big) - B_3 \bigg] \,, \\
 H_3^\Delta
 &=& 2P^+ \, \int dk^- \, \bigg[ \frac{P^2}{2 M^2} \, \big( A_6 - x A_7 \big)
     + \frac{P \cdot k}{M^2} \, A_7 - B_4 \bigg] \,.
\label{e:gtmd_gpcf_16}
\end{eqnarray}

%
%
%
\section{Model calculation of GTMDs}
\label{gtmd_model}
For illustrative purposes and in order to get a first estimate we calculate all 
GTMDs for a spin-0 hadron in a simple spectator model by restricting ourselves to 
lowest nontrivial order in perturbation theory.
To this order only two types of particles have to be considered: the meson (pion) 
target with mass $M$ and quarks/antiquarks with mass $m$. 
In this model a pion, characterized by the field $\varphi$, is coupled to a quark
and an antiquark by means of a pseudo-scalar interaction. 
Including isospin the interaction part of the Lagrangian reads
\begin{equation}
 \mathcal{L}_\text{int}(x) = - i \, g_{\pi} \, \bar{\Psi}(x) \, \gamma_5 \, 
 \vec{\tau} \cdot \vec{\varphi}(x) \, \Psi(x) \,,
\end{equation}
with the coupling constant $g_{\pi}$ and the Pauli matrices $\tau_i$.
The results given below, which are valid for the kinematical range $0 \le x \le 1$, 
represent the valence quark and antiquark GTMDs for neutral pions.
In order to get the valence GTMDs for charged pions one has to multiply
the expressions by a factor two. 
\FIGURE[t]{\includegraphics{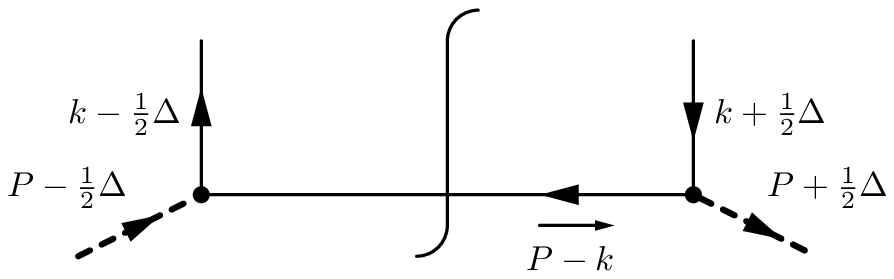}
\caption{Lowest nontrivial order diagram contributing to the GTMDs in the spectator 
model for a pion.}
\label{f:model}}

The lowest order contribution to the generalized $k_T$-dependent correlator 
in eq.~(\ref{e:corr_gtmd}) comes from the tree-level diagram depicted in 
figure~\ref{f:model}.
This diagram can be evaluated in a straightforward manner yielding 
\begin{align} \label{e:corr_model}
& W^{[\Gamma]}(P,x,\vec{k}_T,\Delta,N;\eta) \vphantom{\frac{1}{1}}
\nonumber \\
& \qquad = \frac{g_{\pi}^2 (1-x) \, \text{Tr} \big[ (\Pslash - \kslash + m) \, 
          (\kslash + \tfrac{1}{2} \Dslash + m) \, 
          \Gamma \, (\kslash - \tfrac{1}{2} \Dslash + m) \big]}
          {4(2\pi)^3 \, (1-\xi^2) P^+ \, D_+ \, D_-} \,,
\end{align}
where the denominators $D_\pm$ are given by
\begin{equation}
 D_\pm
 =\bigg( \vec{k}_T \pm \frac{(1-x)}{2(1\mp\xi)} \, \vec{\Delta}_T \bigg)^2
  + m^2 - \frac{(1-x) \, (x\mp\xi)}{(1\mp\xi)^2} \, M^2 \,,
\end{equation}
and $k^-$ is fixed by the cut in the diagram to be
\begin{equation}
 k^-
 =\frac{\tfrac{1}{4} \vec{\Delta}_T^2 + M^2}{2(1-\xi^2)P^+} 
 - \frac{\vec{k}_T^2 + m^2}{2(1-x)P^+} \,.
\end{equation}
Since the calculation is carried out only to lowest order in perturbation theory,
no effect due to the Wilson line enters.
As a consequence, the trace in~(\ref{e:corr_model}) actually does not depend on the 
parameter $\eta$.

Using now the expression~(\ref{e:corr_model}) and the definitions for the GTMDs 
in eqs.~(\ref{e:gtmd_1})--(\ref{e:gtmd_12}) one obtains
\begin{eqnarray}
 E_2^{e}
 &=& \frac{4C \, m}{(1-\xi^2) M} \, \bigg[ \tfrac{1}{4}\vec{\Delta}_T^2 + \tfrac{1}{2}(1+\xi^2) M^2 \bigg] \,, \label{e:gtmd_model_1} \\
 \tilde{E}_2^{e}
 &=& 0 \,,\phantom{\bigg]}\\
 F_1^{e}
 &=& \frac{2C}{1-x} \, \bigg[ (1-\xi^2) \, (\vec{k}_T^2 + m^2) + \xi (1-x) \, \vec{k}_T \cdot \vec{\Delta}_T
     - \tfrac{1}{4}(1-x)^2 \vec{\Delta}_T^2 \bigg] \,,\\
 F_2^{k,e}
 &=& \frac{4C}{1-\xi^2} \, \bigg[ \tfrac{1}{4}\vec{\Delta}_T^2 + \tfrac{1}{2}(1+\xi^2) M^2 \bigg] \,,\\
 F_2^{\Delta,e}
 &=& \frac{C}{(1-x) (1-\xi^2)} \, \bigg[ \xi (1-\xi^2) \, (\vec{k}_T^2 + m^2)
     - (1-x) \, (1-\xi^2) \, \vec{k}_T \cdot \vec{\Delta}_T \nonumber\\*
 & & \hspace{17.8ex} - \xi (1-x)^2 \, (\tfrac{1}{4} \vec{\Delta}_T^2 + M^2) \bigg] \,,\\
 F_3^{e}
 &=& - \frac{C}{(1-x) (1-\xi^2) M^2} \bigg[ \tfrac{1}{4}(1-\xi^2) \, (\vec{k}_T^2 + m^2) \, \vec{\Delta}_T^2
     + \xi (1-x) \, \vec{k}_T \cdot \vec{\Delta}_T \, (\tfrac{1}{4}\vec{\Delta}_T^2 + M^2) \nonumber\\*
 & & \hspace{22.7ex} - (1-x)^2 \, (\tfrac{1}{4}\vec{\Delta}_T^2 + M^2)^2 \bigg]\,,\\
 \tilde{G}_1^{e}
 &=& 2C \, M^2 \,,\phantom{\bigg]}\\
 G_2^{k,e}
 &=& \frac{4C \, \xi}{1-\xi^2} \, \bigg[ \tfrac{1}{4}\vec{\Delta}_T^2 + M^2 \bigg] \,,\\
 G_2^{\Delta,e}
 &=& \frac{C}{(1-x) (1-\xi^2)} \, \bigg[ (1-\xi^2) \, (\vec{k}_T^2 + m^2)
     - (1-x)^2 \, (\tfrac{1}{4}\vec{\Delta}_T^2 + M^2) \bigg] \,,\\
 \tilde{G}_3^{e}
 &=& - \frac{C}{1-\xi^2} \, \bigg[ \tfrac{1}{4}\vec{\Delta}_T^2 + M^2 \bigg] \,,\\
 H_1^{k,e}
 &=& 0 \,,\phantom{\bigg]}
\label{e:gtmd_model_11} \\
 H_1^{\Delta,e}
 &=& -2C \, mM \,,\phantom{\bigg]}
\label{e:gtmd_model_12} \\
 H_2^{e}
 &=& \frac{4C \, \xi m}{(1-\xi^2) M} \, \bigg[ \tfrac{1}{4}\vec{\Delta}_T^2 + M^2 \bigg] \,,\\
 \tilde{H}_2^{e}
 &=& 0 \,,\phantom{\bigg]}\\
 H_3^{k,e}
 &=& 0 \,,\phantom{\bigg]}\\
 H_3^{\Delta,e}
 &=& \frac{C \, m}{(1-\xi^2) M} \, \bigg[ \tfrac{1}{4}\vec{\Delta}_T^2 + M^2 \bigg] \,,
 \label{e:gtmd_model_16}
\end{eqnarray}
with
\begin{equation}
C = \frac{g_{\pi}^2 (1-x)}{2(2\pi)^3 \, (1-\xi^2) \, D_+ \, D_-} \,.
\end{equation}
To shorten the notation we have suppressed the arguments of the GTMDs.
All (na\"ive) T-odd GTMDs vanish to lowest order in perturbation theory 
investigated here.
To get nonzero results for these functions requires considering at least 
one-loop corrections that include effects from the Wilson line.
On the other hand, the vanishing of the T-even parts of $\tilde{E}_2^e$, $H_1^{k,e}$, 
$\tilde{H}_2^e$, $H_3^{k,e}$ has to be attributed to the simplicity of the model.

%
%
%
\bibliographystyle{JHEP}
\bibliography{gtmd_1}

\providecommand{\href}[2]{#2}\begingroup\raggedright\begin{thebibliography}{10}

\bibitem{Mueller:1998fv}
D.~M{\"u}ller, D.~Robaschik, B.~Geyer, F.M.~Dittes and J.~Horejsi, {\it Wave
  functions, evolution equations and evolution kernels from light-ray operators
  of {QCD}},  {\em Fortschr. Phys.} {\bf 42} (1994) 101
  [\href{http://xxx.lanl.gov/abs/hep-ph/9812448}{{\tt hep-ph/9812448}}].

\bibitem{Goeke:2001tz}
K.~Goeke, M.V.~Polyakov and M.~Vanderhaeghen, {\it Hard exclusive reactions
  and the structure of hadrons},  {\em Prog. Part. Nucl. Phys.} {\bf 47} (2001)
  401 [\href{http://xxx.lanl.gov/abs/hep-ph/0106012}{{\tt hep-ph/0106012}}].

\bibitem{Diehl:2003ny}
M.~Diehl, {\it Generalized parton distributions},  {\em Phys. Rept.} {\bf 388}
  (2003) 41 [\href{http://xxx.lanl.gov/abs/hep-ph/0307382}{{\tt
  hep-ph/0307382}}].

\bibitem{Belitsky:2005qn}
A.V. Belitsky and A.V. Radyushkin, {\it Unraveling hadron structure with
  generalized parton distributions},  {\em Phys. Rept.} {\bf 418} (2005) 1
  [\href{http://xxx.lanl.gov/abs/hep-ph/0504030}{{\tt hep-ph/0504030}}].

\bibitem{Boffi:2007yc}
S.~Boffi and B.~Pasquini, {\it Generalized parton distributions and the
  structure of the nucleon},  {\em Riv. Nuovo Cim.} {\bf 30} (2007) 387
  [\href{http://xxx.lanl.gov/abs/0711.2625}{{\tt arXiv:0711.2625}}].

\bibitem{Burkardt:2000za}
M.~Burkardt, {\it Impact parameter dependent parton distributions and
  off-forward parton distributions for $\zeta \to 0$},  {\em Phys. Rev.} {\bf D
  62} (2000) 071503 [{\it Erratum ibid.} {\bf 66} (2002) 119903]
  [\href{http://xxx.lanl.gov/abs/hep-ph/0005108}{{\tt hep-ph/0005108}}].

\bibitem{Ralston:2001xs}
J.P.~Ralston and B.~Pire, {\it Femto-photography of protons to nuclei with
  deeply virtual {Compton} scattering},  {\em Phys. Rev.} {\bf D 66} (2002)
  111501 [\href{http://xxx.lanl.gov/abs/hep-ph/0110075}{{\tt
  hep-ph/0110075}}].

\bibitem{Diehl:2002he}
M.~Diehl, {\it Generalized parton distributions in impact parameter space},
  {\em Eur. Phys. J.} {\bf C 25} (2002) 223 [{\it Erratum ibid.} {\bf 31}
  (2003) 277] [\href{http://xxx.lanl.gov/abs/hep-ph/0205208}{{\tt
  hep-ph/0205208}}].

\bibitem{Burkardt:2002hr}
M.~Burkardt, {\it Impact parameter space interpretation for generalized parton
  distributions},  {\em Int. J. Mod. Phys.} {\bf A 18} (2003) 173
  [\href{http://xxx.lanl.gov/abs/hep-ph/0207047}{{\tt hep-ph/0207047}}].

\bibitem{Mulders:1995dh}
P.J.~Mulders and R.D.~Tangerman, {\it The complete tree-level result up to
  order $1/{Q}$ for polarized deep-inelastic leptoproduction},  {\em Nucl.
  Phys.} {\bf B 461} (1996) 197 [{\it Erratum ibid.} {\bf 484} (1997) 538]
  [\href{http://xxx.lanl.gov/abs/hep-ph/9510301}{{\tt hep-ph/9510301}}].

\bibitem{Bacchetta:2006tn}
A.~Bacchetta {\em et~al.}, {\it Semi-inclusive deep inelastic scattering at
  small transverse momentum},  {\em JHEP} {\bf 02} (2007) 093
  [\href{http://xxx.lanl.gov/abs/hep-ph/0611265}{{\tt hep-ph/0611265}}].

\bibitem{Barone:2001sp}
V.~Barone, A.~Drago and P.G.~Ratcliffe, {\it Transverse polarisation of
  quarks in hadrons},  {\em Phys. Rept.} {\bf 359} (2002) 1
  [\href{http://xxx.lanl.gov/abs/hep-ph/0104283}{{\tt hep-ph/0104283}}].

\bibitem{D'Alesio:2007jt}
U.~D'Alesio and F.~Murgia, {\it Azimuthal and single spin asymmetries in hard
  scattering processes},  \href{http://xxx.lanl.gov/abs/0712.4328}{{\tt
  arXiv:0712.4328}}.

\bibitem{Burkardt:2002ks}
M.~Burkardt, {\it Impact parameter dependent parton distributions and
  transverse single spin asymmetries},  {\em Phys. Rev.} {\bf D 66} (2002)
  114005 [\href{http://xxx.lanl.gov/abs/hep-ph/0209179}{{\tt
  hep-ph/0209179}}].

\bibitem{Burkardt:2003uw}
M.~Burkardt, {\it Chromodynamic lensing and transverse single spin
  asymmetries},  {\em Nucl. Phys.} {\bf A 735} (2004) 185
  [\href{http://xxx.lanl.gov/abs/hep-ph/0302144}{{\tt hep-ph/0302144}}].

\bibitem{Burkardt:2003je}
M.~Burkardt and D.S.~Hwang, {\it Sivers asymmetry and generalized parton
  distributions in impact parameter space},  {\em Phys. Rev.} {\bf D 69} (2004)
  074032 [\href{http://xxx.lanl.gov/abs/hep-ph/0309072}{{\tt
  hep-ph/0309072}}].

\bibitem{Diehl:2005jf}
M.~Diehl and P.~H{\"a}gler, {\it Spin densities in the transverse plane and
  generalized transversity distributions},  {\em Eur. Phys. J.} {\bf C 44}
  (2005) 87 [\href{http://xxx.lanl.gov/abs/hep-ph/0504175}{{\tt
  hep-ph/0504175}}].

\bibitem{Burkardt:2005hp}
M.~Burkardt, {\it Transverse deformation of parton distributions and
  transversity decomposition of angular momentum},  {\em Phys. Rev.} {\bf D 72}
  (2005) 094020 [\href{http://xxx.lanl.gov/abs/hep-ph/0505189}{{\tt
  hep-ph/0505189}}].

\bibitem{Lu:2006kt}
Z.~Lu and I.~Schmidt, {\it Connection between the {Sivers} function and the
  anomalous magnetic moment},  {\em Phys. Rev.} {\bf D 75} (2007) 073008
  [\href{http://xxx.lanl.gov/abs/hep-ph/0611158}{{\tt hep-ph/0611158}}].

\bibitem{Meissner:2007rx}
S.~Meissner, A.~Metz and K.~Goeke, {\it Relations between generalized and
  transverse momentum dependent parton distributions},  {\em Phys. Rev.} {\bf
  D 76} (2007) 034002 [\href{http://xxx.lanl.gov/abs/hep-ph/0703176}{{\tt
  hep-ph/0703176}}].

\bibitem{Sivers:1989cc}
D.W.~Sivers, {\it Single spin production asymmetries from the hard scattering
  of point-like constituents},  {\em Phys. Rev.} {\bf D 41} (1990) 83.

\bibitem{Sivers:1990fh}
D.W.~Sivers, {\it Hard scattering scaling laws for single spin production
  asymmetries},  {\em Phys. Rev.} {\bf D 43} (1991) 261.

\bibitem{Boer:1997nt}
D.~Boer and P.J.~Mulders, {\it Time-reversal odd distribution functions in
  leptoproduction},  {\em Phys. Rev.} {\bf D 57} (1998) 5780
  [\href{http://xxx.lanl.gov/abs/hep-ph/9711485}{{\tt hep-ph/9711485}}].

\bibitem{Ji:2003ak}
X.-d. Ji, {\it Viewing the proton through 'color'-filters},  {\em Phys. Rev.
  Lett.} {\bf 91} (2003) 062001
  [\href{http://xxx.lanl.gov/abs/hep-ph/0304037}{{\tt hep-ph/0304037}}].

\bibitem{Belitsky:2003nz}
A.V.~Belitsky, X.-d.~Ji and F.~Yuan, {\it Quark imaging in the proton via
  quantum phase-space distributions},  {\em Phys. Rev.} {\bf D 69} (2004)
  074014 [\href{http://xxx.lanl.gov/abs/hep-ph/0307383}{{\tt
  hep-ph/0307383}}].

\bibitem{Meissner:2007ez}
S.~Meissner, A.~Metz, M.~Schlegel and K.~Goeke, {\it Relations between {GPDs}
  and {TMDs}: model results and beyond},
  \href{http://xxx.lanl.gov/abs/0710.5846}{{\tt arXiv:0710.5846}}.

\bibitem{Vanderhaeghen:1999xj}
M.~Vanderhaeghen, P.A.M.~Guichon and M.~Guidal, {\it Deeply virtual
  electroproduction of photons and mesons on the nucleon: leading order
  amplitudes and power corrections},  {\em Phys. Rev.} {\bf D 60} (1999)
  094017 [\href{http://xxx.lanl.gov/abs/hep-ph/9905372}{{\tt
  hep-ph/9905372}}].

\bibitem{Diehl:2007hd}
M.~Diehl and W.~Kugler, {\it Next-to-leading order corrections in exclusive
  meson production},  {\em Eur. Phys. J.} {\bf C 52} (2007) 933
  [\href{http://xxx.lanl.gov/abs/0708.1121}{{\tt arXiv:0708.1121}}].

\bibitem{Goloskokov:2007nt}
S.V.~Goloskokov and P.~Kroll, {\it The role of the quark and gluon {GPDs} in
  hard vector-meson electroproduction},  {\em Eur. Phys. J.} {\bf C 53} (2008)
  367 [\href{http://xxx.lanl.gov/abs/0708.3569}{{\tt arXiv:0708.3569}}].

\bibitem{Collins:2007ph}
J.C.~Collins, T.C.~Rogers and A.M.~Stasto, {\it Fully unintegrated parton
  correlation functions and factorization in lowest order hard scattering},
  {\em Phys. Rev.} {\bf D 77} (2008) 085009
  [\href{http://xxx.lanl.gov/abs/0708.2833}{{\tt arXiv:0708.2833}}].

\bibitem{Martin:1999wb}
A.D.~Martin, M.G.~Ryskin and T.~Teubner, {\it {$Q^2$} dependence of
  diffractive vector meson electroproduction},  {\em Phys. Rev.} {\bf D 62}
  (2000) 014022 [\href{http://xxx.lanl.gov/abs/hep-ph/9912551}{{\tt
  hep-ph/9912551}}].

\bibitem{Khoze:2000cy}
V.A.~Khoze, A.D.~Martin and M.G.~Ryskin, {\it Can the {Higgs} be seen in
  rapidity gap events at the {Tevatron} or the {LHC?}},  {\em Eur. Phys. J.}
  {\bf C 14} (2000) 525 [\href{http://xxx.lanl.gov/abs/hep-ph/0002072}{{\tt
  hep-ph/0002072}}].

\bibitem{Goeke:2003az}
K.~Goeke, A.~Metz, P.V.~Pobylitsa and M.V.~Polyakov, {\it Lorentz invariance
  relations among parton distributions revisited},  {\em Phys. Lett.} {\bf B
  567} (2003) 27 [\href{http://xxx.lanl.gov/abs/hep-ph/0302028}{{\tt
  hep-ph/0302028}}].

\bibitem{Bacchetta:2004zf}
A.~Bacchetta, P.J.~Mulders and F.~Pijlman, {\it New observables in
  longitudinal single-spin asymmetries in semi-inclusive {DIS}},  {\em Phys.
  Lett.} {\bf B 595} (2004) 309
  [\href{http://xxx.lanl.gov/abs/hep-ph/0405154}{{\tt hep-ph/0405154}}].

\bibitem{Goeke:2005hb}
K.~Goeke, A.~Metz and M.~Schlegel, {\it Parameterization of the quark-quark
  correlator of a spin-1/2 hadron},  {\em Phys. Lett.} {\bf B 618} (2005) 90
  [\href{http://xxx.lanl.gov/abs/hep-ph/0504130}{{\tt hep-ph/0504130}}].

\bibitem{Collins:2002kn}
J.C.~Collins, {\it Leading-twist single-transverse-spin asymmetries:
  {Drell-Yan} and deep-inelastic scattering},  {\em Phys. Lett.} {\bf B 536}
  (2002) 43 [\href{http://xxx.lanl.gov/abs/hep-ph/0204004}{{\tt
  hep-ph/0204004}}].

\bibitem{Collins:1981uw}
J.C.~Collins and D.E.~Soper, {\it Parton distribution and decay functions},
  {\em Nucl. Phys.} {\bf B 194} (1982) 445.

\bibitem{Collins:1999dz}
J.C.~Collins and F.~Hautmann, {\it Infrared divergences and non-lightlike
  eikonal lines in {Sudakov} processes},  {\em Phys. Lett.} {\bf B 472} (2000)
  129 [\href{http://xxx.lanl.gov/abs/hep-ph/9908467}{{\tt hep-ph/9908467}}].

\bibitem{Collins:2000gd}
J.C.~Collins and F.~Hautmann, {\it Soft gluons and gauge-invariant
  subtractions in {NLO} parton-shower {Monte Carlo} event generators},  {\em
  JHEP} {\bf 03} (2001) 016
  [\href{http://xxx.lanl.gov/abs/hep-ph/0009286}{{\tt hep-ph/0009286}}].

\bibitem{Ji:2002aa}
X.-d.~Ji and F.~Yuan, {\it Parton distributions in light-cone gauge: Where are
  the final-state interactions?},  {\em Phys. Lett.} {\bf B 543} (2002) 66
  [\href{http://xxx.lanl.gov/abs/hep-ph/0206057}{{\tt hep-ph/0206057}}].

\bibitem{Belitsky:2002sm}
A.V.~Belitsky, X.-d.~Ji and F.~Yuan, {\it Final state interactions and gauge
  invariant parton distributions},  {\em Nucl. Phys.} {\bf B 656} (2003) 165
  [\href{http://xxx.lanl.gov/abs/hep-ph/0208038}{{\tt hep-ph/0208038}}].

\bibitem{Collins:2004nx}
J.C.~Collins and A.~Metz, {\it Universality of soft and collinear factors in
  hard-scattering factorization},  {\em Phys. Rev. Lett.} {\bf 93} (2004)
  252001 [\href{http://xxx.lanl.gov/abs/hep-ph/0408249}{{\tt
  hep-ph/0408249}}].

\bibitem{Cherednikov:2007tw}
I.O.~Cherednikov and N.G.~Stefanis, {\it Renormalization, {Wilson} lines, and
  transverse-momentum dependent parton distribution functions},  {\em Phys.
  Rev.} {\bf D 77} (2008) 094001
  [\href{http://xxx.lanl.gov/abs/0710.1955}{{\tt arXiv:0710.1955}}].

\bibitem{Cherednikov:2008ua}
I.O.~Cherednikov and N.G.~Stefanis, {\it Wilson lines and transverse-momentum
  dependent parton distribution functions: A renormalization-group analysis},
  {\em Nucl. Phys.} {\bf B 802} (2008) 146
  [\href{http://xxx.lanl.gov/abs/0802.2821}{{\tt arXiv:0802.2821}}].

\bibitem{Collins:2003fm}
J.C.~Collins, {\it What exactly is a parton density?},  {\em Acta Phys.
  Polon.} {\bf B 34} (2003) 3103
  [\href{http://xxx.lanl.gov/abs/hep-ph/0304122}{{\tt hep-ph/0304122}}].

\bibitem{Hautmann:2007uw}
F.~Hautmann, {\it Endpoint singularities in unintegrated parton distributions},
   {\em Phys. Lett.} {\bf B 655} (2007) 26
  [\href{http://xxx.lanl.gov/abs/hep-ph/0702196}{{\tt hep-ph/0702196}}].

\bibitem{Anikin:2000em}
I.V.~Anikin, B.~Pire and O.V.~Teryaev, {\it On the gauge invariance of the
  {DVCS} amplitude},  {\em Phys. Rev.} {\bf D 62} (2000) 071501
  [\href{http://xxx.lanl.gov/abs/hep-ph/0003203}{{\tt hep-ph/0003203}}].

\bibitem{Burkardt:2007xm}
M.~Burkardt and B.~Hannafious, {\it Are all {Boer-Mulders} functions alike?},
  {\em Phys. Lett.} {\bf B 658} (2008) 130
  [\href{http://xxx.lanl.gov/abs/0705.1573}{{\tt arXiv:0705.1573}}].

\bibitem{Diehl:1998pd}
M.~Diehl, T.~Gousset and B.~Pire, {\it Exclusive electroproduction of vector
  mesons and transversity distributions},  {\em Phys. Rev.} {\bf D 59} (1999)
  034023 [\href{http://xxx.lanl.gov/abs/hep-ph/9808479}{{\tt
  hep-ph/9808479}}].

\bibitem{Collins:1999un}
J.C.~Collins and M.~Diehl, {\it Transversity distribution does not contribute
  to hard exclusive electroproduction of mesons},  {\em Phys. Rev.} {\bf D 61}
  (2000) 114015 [\href{http://xxx.lanl.gov/abs/hep-ph/9907498}{{\tt
  hep-ph/9907498}}].

\bibitem{Ivanov:2002jj}
D.Y.~Ivanov, B.~Pire, L.~Szymanowski and O.V.~Teryaev, {\it Probing
  chiral-odd {GPDs} in diffractive electroproduction of two vector mesons},
  {\em Phys. Lett.} {\bf B 550} (2002) 65
  [\href{http://xxx.lanl.gov/abs/hep-ph/0209300}{{\tt hep-ph/0209300}}].

\bibitem{Brommel:2007xd}
{\bf QCDSF and UKQCD} Collaboration, D.~Br{\"o}mmel {\em et~al.}, {\it The
  spin structure of the pion},  \href{http://xxx.lanl.gov/abs/0708.2249}{{\tt
  arXiv:0708.2249}}.

\bibitem{Goldstein:2002vv}
G.R.~Goldstein and L.P.~Gamberg, {\it Transversity and meson
  photoproduction},  \href{http://xxx.lanl.gov/abs/hep-ph/0209085}{{\tt
  hep-ph/0209085}}.

\bibitem{Boer:2002ju}
D.~Boer, S.J.~Brodsky and D.S.~Hwang, {\it Initial state interactions in the
  unpolarized {Drell-Yan} process},  {\em Phys. Rev.} {\bf D 67} (2003) 054003
  [\href{http://xxx.lanl.gov/abs/hep-ph/0211110}{{\tt hep-ph/0211110}}].

\bibitem{Gamberg:2003ey}
L.P.~Gamberg, G.R.~Goldstein and K.A.~Oganessyan, {\it Novel transversity
  properties in semi-inclusive deep inelastic scattering},  {\em Phys. Rev.}
  {\bf D 67} (2003) 071504 [\href{http://xxx.lanl.gov/abs/hep-ph/0301018}{{\tt
  hep-ph/0301018}}].

\bibitem{Bacchetta:2003rz}
A.~Bacchetta, A.~Schaefer and J.-J.~Yang, {\it Sivers function in a spectator
  model with axial-vector diquarks},  {\em Phys. Lett.} {\bf B 578} (2004) 109
  [\href{http://xxx.lanl.gov/abs/hep-ph/0309246}{{\tt hep-ph/0309246}}].

\bibitem{Gamberg:2007wm}
L.P.~Gamberg, G.R.~Goldstein and M.~Schlegel, {\it Transverse quark spin
  effects and the flavor dependence of the {Boer-Mulders} function},  {\em
  Phys. Rev.} {\bf D 77} (2008) 094016
  [\href{http://xxx.lanl.gov/abs/0708.0324}{{\tt arXiv:0708.0324}}].

\bibitem{Lu:2004hu}
Z.~Lu and B.-Q.~Ma, {\it Non-zero transversity distribution of the pion in a
  quark-spectator-antiquark model},  {\em Phys. Rev.} {\bf D 70} (2004) 094044
  [\href{http://xxx.lanl.gov/abs/hep-ph/0411043}{{\tt hep-ph/0411043}}].

\end{thebibliography}\endgroup
\end{document}